\documentclass[12pt,a4paper]{article}
 \usepackage{epsf}
 \usepackage[dvips]{graphicx}
\begin{document}
\textwidth=135mm
\textheight=200mm

 \begin{center}
  {\bfseries Delayed clusters accompanying nonmesonic weak decay
  of the $\Lambda$-hypernuclei:
  a clue to nonleptonic processes }
  \vskip 5mm
  L.~Majling$^a$ and  O.~Majlingov\'a$^b$
  \vskip 5mm
  {\small {\it $^a$ Nuclear Physics Institute,
  Academy of Sciences, 
  \v{R}e\v{z}, Czech Republic}}
  \\
  {\small {\it $^b$
   Dept. of Mathematic,
   Czech Technical University, Prague, Czech Republic
  }}
  \\
 \end{center}
 \vskip 5mm
 \centerline{\bf Abstract}
 The nonmesonic decay of $\Lambda$-hypernuclei
 provides access to the nonleptonic weak decay
 process $\Lambda N \to NN$, which is achievable only through the
 observation  of hypernuclear ground-state decays.
 We continue the discussion of some specific cases which make it
 possible  to detect a few exclusive transitions,
 namely, the stripping of nucleon from the ground state
 results in a resonance state decaying via emission of two
 clusters.
 Delayed clusters accompanying weak decay of light hypernuclei
 give a unique information on spin dependence of the
 weak decay matrix elements. \\[2mm]
 \vskip 10mm
 \section{Nonmesonic decay of $\mathbf{\Lambda}$-hypernuclei}
 $\Lambda$-hypernuclei,  bound systems with nucleons and one $\Lambda$-hyperon,
 baryon with a new flavor - strangeness, are not only exotic (the third dimension
 in the chart of nuclei, new symmetries, {\it  etc.}) \cite{Greiner}.
 The realization  of the brilliant suggestion by Podgoretskii \cite{Pod63}
 to study hypernuclear production in the strangeness exchange reaction
 $(K^-,\pi^-)$ under proper kinematic conditions,
 opens a way to study spectra of several dozen hypernuclei \cite{HT06}.
 They provide an excellent tool for obtaining information on
 $\Lambda N$ interaction and explore the full SU(3) symmetry
 breaking baryon-baryon interaction.
 The non mesonic weak decay (NMWD) is the only way available to
 study the strangeness-changing interaction between baryons.
 The strangeness changing process in which a $\Lambda$-hyperon
 converts into a neutron with the release of up to 176 MeV
 provides a clear signal for conversion of an
 $s$-quark to a $u$- or $d$-quark.
 The weak interaction at the quark level is shortranged,
 involving $W$-, $Z$-exchange \cite{Ok81}.
 Baryon-baryon interaction is modelled in terms of
 one-meson-exchange interaction, all pseudoscalar $(\pi, \eta, K)$
 and  vector $(\rho, \omega, K^*)$ meson exchanges included.
 The evaluation of the weak baryon-baryon-meson vertices
 is quark model based but presented in terms of the symmetry
 $SU(6)_w$ \cite{Par}.
 Recently, quality of experimental data on NMWD has improved
 considerably \cite{Bhang}.

\newpage
 \section{Delayed clusters }
\begin{figure}[h]
  \centering
  \includegraphics*[height=81mm,angle=0]{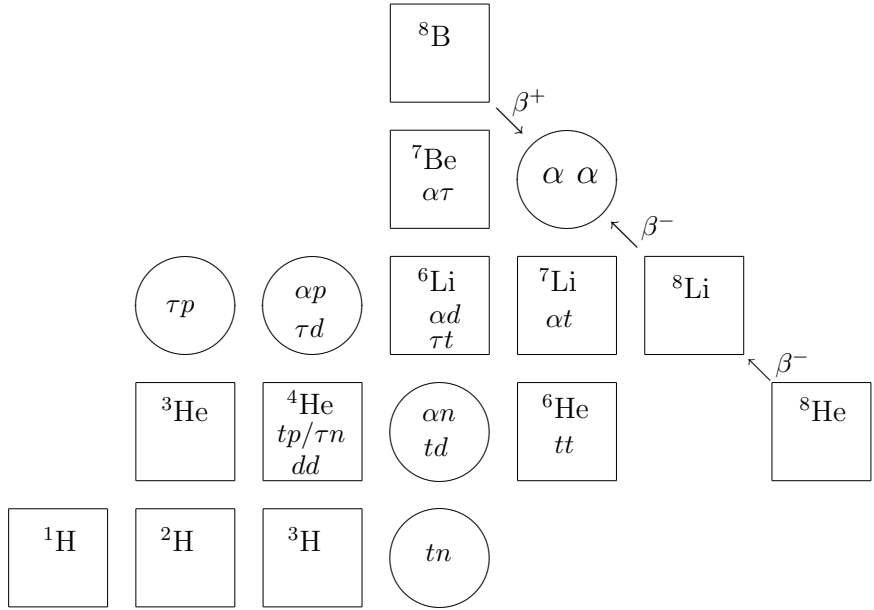}
  \caption{Light nuclei and their cluster structure.
  The circle denotes unstable species.}
  \label{fig:chart}
 \end{figure}

 We explore the well-known fact that in several light
 $p$-shell nuclei the stripping of nucleon from the ground state
 (the one-nucleon induced non-mesonic weak decay mechanism) \cite{SP02}
 results in a resonance state decaying via emission of two
 light nuclei (clusters):
 \begin{equation}
  \label{eq:ClDec}
  \begin{array}{lclcccl}
   {^A_\Lambda}Z \rightarrow (n + N) & + &
   {^{A_f}Z^*_f} (E; \; J^{\pi} T) & \\
   & & & & {^{A_1}Z_1} & \\[-3mm]
   & & & \nearrow & \\[-3mm]
   & & {^{A_f}Z^*_f} (E; \; J^{\pi} T) & & & E \Rightarrow J^{\pi} T\\[-3mm]
   & & & \searrow & \\[-3mm]
   & & & & {^{A_2}Z_2} & \\
  \end{array}
 \end{equation}
 The energy $E$ between clusters  ${^{A_1}{\rm Z}_1}$ and
 ${^{A_2}{\rm Z}_2}$ determines the quantum numbers   $(J^{\pi}T)$
 of the resonance state in  ${^{A_f}{\rm Z}_f}$,
 so in this a case it is possible to study  the {\it exclusive channel }
 of NMWD.
 The most familiar is  the removal of a neutron from
 $^9$Be which leads to the formation of $^8$Be nucleus in several states
 emitting two $\alpha$ particles  \cite{MB00}.
 Now, we look for other examples among $p$-shell hypernuclei
 \cite{OM08}.
 The proper candidates for delayed cluster decay are seen in
 Fig.~\ref{fig:chart}.

\newpage
 We analyze not only the decays  with $\alpha$ particles
 ($\Lambda$ strips the nucleon from $p$-shell)
 but also the decays with three-nucleon  clusters ($t, \tau$) -
 $\Lambda$ strips the nucleon from $s$-shell \cite{Ba08}.
 \begin{equation}
  \label{eq:HYP}
  \begin{array}{lclclccl}
   & & & & & &\multicolumn{2}{l}{\alpha + "(k-1)"}\\[-3mm]
   & & & & & \multicolumn{1}{r}{\nearrow}& \\[-3mm]
   & & {s_\Lambda\ p} & +& s^4\ p^{k-1} & \\[-3mm]
   & \nearrow & \\[-3mm]
   s^4\ p^k\ s_\Lambda & \\[-3mm]
   & \searrow & \\[-3mm]
   & & {s_\Lambda\ s} & + & s^3\ p^k & \\[-3mm]
   & & & & & \multicolumn{1}{r}{\searrow}& \\[-3mm]
   & & & & & &\multicolumn{2}{l}{ 3N + "k"}
   \\
  \end{array}
 \end{equation}

 \begin{table}[h]
  \centering
  \caption{\bf
  Possible clusters accompanying one-nucleon induced ($\Gamma_{1N}$)
  and two-nucleon induced ($\Gamma_{2N}$)  decay of hypernuclei}
  \vspace{-5mm}
  \label{ac}
  $$
  \begin{array}{|l|c|c|c|c|c|}
   \hline
   &\multicolumn{2}{|c|}{ }& \multicolumn{3}{|c|}{ }
   \\[-4mm]
   &\multicolumn{2}{|c|}{\Gamma_{1N}}&
   \multicolumn{3}{|c|}{\Gamma_{2N}}\\[.5mm]
   \cline{2-6}
   & & & & & \\[-4mm]
   {^A_\Lambda}{\rm Z }
   &\Gamma_n&\Gamma_p&\Gamma_{nn}&\Gamma_{np}&\Gamma_{pp}
   \\[.5mm]
   \hline
   & & & & & \\[-4mm]
   {^7_\Lambda}{\rm He}
   &\alpha n + t d &  &
   d d + t p + \tau n & t n &
   \\[1mm]
   {^7_\Lambda}{\rm Li}
   & \alpha p + \tau d & \alpha n + d t & \tau p &
   dd  + t p + \tau n  & t n
   \\[1mm]
   {^7_\Lambda}{\rm Be}
   & &\alpha p + \tau d &  & \tau p &
   dd  + t p + \tau n
   \\[2mm]
   {^8_\Lambda}{\rm He}  & tt & &  \alpha n + t d & &
   \\[1mm]
   {^8_\Lambda}{\rm Li}
   & \alpha d +  \tau t & t t &  \alpha p + \tau d &\alpha n + t d &
   \\[1mm]
   {^8_\Lambda}{\rm Be}
   & & \alpha d + \tau t& & \alpha p + \tau d& \alpha n +  t d
   \\[1mm]
   {^8_\Lambda}{\rm B}
   & & & &  & \alpha p + \tau d
   \\[2mm]
   {^9_\Lambda}{\rm Li}
   & \alpha t &&\alpha d + \tau t & t t &
   \\[1mm]
   {^9_\Lambda}{\rm Be}
   &\alpha \tau  & \alpha t & \tau \tau &\alpha d +\tau t & t t
   \\[1mm]
   {^9_\Lambda}{\rm B}
   & & \alpha \tau & &\tau \tau & \alpha d + \tau t
   \\[2mm]
   {^{10}_{\; \Lambda}}{\rm Li}
   & ^8{\rm Li} & ^8{\rm He} &  
   \alpha t & &
   \\[1mm]
   {^{10}_{\; \Lambda}}{\rm Be}
   & \alpha \alpha & ^8{\rm Li}  & \alpha \tau & \alpha t &
   \\[1mm]
   {^{10}_{\; \Lambda}}{\rm B}
   &^8{\rm B}  & \alpha \alpha & & \alpha \tau & \alpha t
   \\[1mm]
   {^{10}_{\; \Lambda}}{\rm C}
   & & ^8{\rm B}  & & & \alpha \tau
   \\[2mm]
   {^{11}_{\; \Lambda}}{\rm Be}
   & & & \alpha \alpha & ^8{\rm Li} & ^8{\rm He}
   \\[1mm]
   {^{11}_{\; \Lambda}}{\rm B}
   & & & ^8{\rm B} &  \alpha \alpha & ^8{\rm Li}
   \\[1mm]
   {^{11}_{\; \Lambda}}{\rm C}
   & & & & ^8{\rm B} &
   \alpha \alpha
   \\[1mm]
   \hline
  \end{array}
  $$
 \end{table}

 The population of final states is governed by spectroscopic factors.
 They are basic  nuclear structure ingredients in transition amplitudes
 for direct nuclear reactions.
 When $\Lambda$-hyperon interacts with valence nucleon, $\alpha$ clusters appear.
 Tables of fractional parentage coefficients (FPC) are a standard part of the shell model
 \cite{NS}.
 The emission of 3N clusters in the two-body decay
 requests more sophisticated methods of computing FPC for separation
 of the nucleon from the $s$-shell;
 we explored Translational Invariant Shell Model \cite{KSM}.

 \section{Phenomenological weak $\mathbf{\Lambda}$N interaction}
 In the first phenomenological analysis,
 Block and Dalitz (BD)  \cite{BD63}
 expressed the total NM width as a sum of four rates $R_{N S}$
 ($\rho_A$ is the nucleon density in the  hypernucleus).
 The different spin - isospin structure of the ground states of
 four $s$-shell hypernuclei leads to four equations:
 \begin{equation}
  \label{eq:BD}
  \begin{tabular}{l  l }
   $\Gamma_{\rm nm}$($^3_\Lambda$H)&
   = $\varrho_3 /8$  $\cdot$
   (3 R$_{n0}$ + 1 R$_{n1}$ + 3 R$_{p0}$ + 1 R$_{p1}$)\\[1mm]
   $\Gamma_{\rm nm}$($^4_\Lambda$H) \
   $\equiv \; \Gamma^n_{\rm H} + \Gamma^p_{\rm H}$&
   = $\varrho_4 /6$  $\cdot$
   (1 R$_{n0}$ + 3 R$_{n1}$ + 2 R$_{p0}$ + 0 R$_{p1}$)\\[1mm]
   $\Gamma_{\rm nm}$($^4_\Lambda$He)
   $\equiv \Gamma^n_{\rm He} + \Gamma^p_{\rm He}$&
   = $\varrho_4 /6$ $\cdot$
   (2 R$_{n0}$ + 0 R$_{n1}$ + 1 R$_{p0}$ + 3 R$_{p1}$)\\[1mm]
   $\Gamma_{\rm nm}$($^5_\Lambda$He)&
   = $\varrho_5 /8$  $\cdot$
   (1 R$_{n0}$ + 3 R$_{n1}$ + 1 R$_{p0}$ + 3 R$_{p1}$)\\
  \end{tabular}
 \end{equation}
 These relations have an appealing simple form.
 However, it is still impossible to solve
 a set of four equations (\ref{eq:BD}),
 since there are no input data on neutron transitions.
 Nevertheless, BD analysis has so far been a starting point
 in discussing weak decay mechanisms \cite{Sch}.

 The exclusive 3N cluster decay widths for $^7_\Lambda$Li,
 (Eq.~\ref{eq:Li}),
 are determined by the interaction of $\Lambda$ with $s$-shell nucleons,
 see (Eq.~\ref{eq:HYP}),
 so, we could use them in a  phenomenological analysis.
 \begin{equation}
  \label{eq:Li}
  \begin{tabular}{l  l }
   $\Gamma_{\tau d :\ 3/2}$  = &
   $\rho_7 \cdot \kappa(\frac{3}{2})$ $\cdot$  \ 1 R$_{n1}$
   \\[1mm]
   $\Gamma_{\tau d :\ 1/2}$  = &
   $\rho_7 \cdot \kappa(\frac{1}{2})$ $\cdot$
   ($\frac{1}{4}$  R$_{n1}$  +  $\frac{3}{4}$ R$_{n0}$)
   \\[1mm]
   $\Gamma_{t d :\ 3/2}$   = &
   $\rho_7 \cdot \kappa(\frac{3}{2})$ $\cdot$ \ 1 R$_{p1}$
   \\[1mm]
   $\Gamma_{t d :\ 1/2}$   = &
   $\rho_7 \cdot \kappa(\frac{1}{2})$ $\cdot$
   ($\frac{1}{4}$  R$_{p1}$  +  $\frac{3}{4}$ R$_{p0}$)
   \\[1mm]
  \end{tabular}
 \end{equation}
 Here, $\kappa$(J) are square of spin-isospin FCP. In the most simple case
 \linebreak
 ($^6$Li ground state wave function is $|\ s^4\ p^2 : {^{13}{\rm S}}_1 >$) { }
 $\kappa(\frac{3}{2})=\frac{4}{5}$ and $\kappa(\frac{1}{2}) = \frac{1}{5}$.\\
 In the NMWD of the $^7_\Lambda$Li hypernucleus, a neutron (proton) induced
 process is linked with the escape of charged cluster -  $^3$He ($^3$H),
 and there are two spins ($\frac{3}{2}^+$,  $\frac{1}{2}^+$)
 in residual nuclei.
 One can determine unambiguously all four matrix elements R$_{NS}$.
 The ratios $\mathcal{R}_i$
 \begin{equation}
  \label{eq:rates}
  \begin{tabular}{c c c  }
   $\mathcal{R}_1$ $\equiv$
   {\large $\frac{\Gamma_{\tau d :\ 1/2}}{\Gamma_{\tau d :\
   3/2}}$};
   $\qquad$ &
   $\mathcal{R}_2$ $\equiv$
   {\large $\frac{\Gamma_{t d :\ 1/2}}{\Gamma_{t d :\ 3/2}}$};
   $\qquad$ &
   $\mathcal{R}_3$ $\equiv$
  {\large $\frac{\Gamma_{\tau d :\ 3/2}}{\Gamma_{t d :\ 3/2}}$}
  \\
 \end{tabular}
 \end{equation}
 are almost independent on nuclear structure, so, they could
 discriminate between different models of weak interaction.

 The calculations of NMWD for $s$-shell hypernuclei
 $^4_\Lambda$H and $^4_\Lambda$He
 with various models of weak interaction:
 including two pions (TPE) \cite{IUM}, direct quark (DQ)  \cite{SIO05},
 one meson exchange  (OME) \cite{BGHKP}
 were published recently.

 \begin{table}[h]
  \centering
  \caption{\bf
  Phenomenological interaction}
  \vspace{2mm}
  \label{WI}
  \begin{tabular}{|l|r|r|l|c|r|r|r|}
   \hline
   & & & & &\multicolumn{3}{|c|}{ } \\[-4mm]
   & & \multicolumn{1}{|c|}{$^4_\Lambda$H}&
   \multicolumn{1}{|c|}{$^4_\Lambda$He} &  &    
   \multicolumn{3}{c|}{$^7_\Lambda$Li}
   \\
   \cline{1-4} \cline{6-8}
   & & & & & \multicolumn{3}{|c|}{ } \\[-4mm]
   Ref. & \multicolumn{1}{c|}{model }
   & $\Gamma _n$/$\Gamma _p$ & $\Gamma _n$/$\Gamma _p$ & &
   $\kappa \mathcal{R}_1$ & $\kappa \mathcal{R}_2$ & $\mathcal{R}_3$
   \\
   \cline{1-4} \cline{6-8}
   & & & & & & & \\[-4mm]
   \cite{IUM} &
   $\pi$  &
   4.1192 &  0.0475  & & 3.890  &  1.108  & {\bf 0.075}
   \\
   &+2$\pi / \rho$    &
   9.2497 & 0.0452 & & 2.090  &  {\bf 1.102}  & 0.188
   \\
   &
   +2$\pi / \sigma +\omega$ &
   2.7243 & 0.1302 &  & 6.238  &  1.302  & 0.116
   \\
  & +$\rho$ &
  2.1709 & 0.3631  & & {\bf 8.719}  &  1.896  & 0.233\\
  \cline{1-4} \cline{6-8} 
  & & & & & & &  \\[-4mm]
  \cite{SIO05} &
  ME &
  2.705{ }  &  0.417 & & 6.308  &  2.068  & 0.397
  \\
  & DQ+ &
  0.693{ } & 0.269 & & 4.600 & {\bf 5.500} & {\bf 0.500}
  \\
  \cline{1-4} \cline{6-8} 
  & & & & & & &  \\[-4mm]
  \cite{BGHKP} &
  PSVE &
  9.98$\quad$&  0.062  & &   2.007  &  1.138  & 0.284
  \\
  & PKE  &
  27.9 { } { } &  0.031 & &  {\bf 1.360}  &  1.063  & 0.372
  \\
  & SPKE &
  2.70{ } { } & 0.068 &  &1.831  &  1.127  & 0.368 \\
  \hline
 \end{tabular}
\end{table}
Here, $\kappa \equiv \kappa(\frac{3}{2})/ \kappa(\frac{1}{2})$.

\section{Conclusions}
 Delayed cluster widths $\Gamma_{\tau d : J}$ and $\Gamma_{t d : J}$
 are very sensitive to the model of the weak interaction through the chain\\[1mm]
 \begin{tabular}{lcccccc}
  $\quad \Gamma_{\tau d : J}$
   & & R$_{n \, S}$ & &
   $\Gamma^{\,\rm n}( ^4{\rm H})$, {} $\Gamma^{\,\rm n}( ^4{\rm He})$
   & &   model WI :
   \\[-1mm]
   & $\Leftrightarrow$ & & $\Leftrightarrow$ & & $\Leftrightarrow$ &
   \\[-1mm]
  $\quad \Gamma_{t d : J}$
   & & R$_{p \, S}$ & &
   $\Gamma^{\,\rm p}( ^4{\rm H})$, {} $\Gamma^{\,\rm p}( ^4{\rm He})$
   & &  OME or TPE or  HQ \hspace{3mm} \\[1mm]
   \end{tabular}

  Similar relations were found in \cite{MB00} for $\alpha$-particles
  accompanying the weak decay of $^{10}_{\ \Lambda}$B and
  $^{10}_{\ \Lambda}$Be.
  The results which are expected from
  Nuclotron  \cite{Aver} will be of great value.\\

 {\bf Acknowledgements.}
 The authors thank Yu.~Batusov, R.~Jolos, V.~Kuz'min, J.~Lukstins
 and A.~Parfenov for useful discussion and comments.\\
 Work of L.M. is supported by grant 2002/08/0984 of
 the Grant Agency of the Czech Republic.
 Work of O.M. is supported by Project LA08002 of
 the Ministry of Education, Youth and Sport of the Czech Republic.

\end{document}